\documentclass[twocolumn,showpacs,preprintnumbers,amsmath,amssymb]{revtex4}
\usepackage{graphicx}% Include figure files
\usepackage{dcolumn}% Align table columns on decimal point
\usepackage{bm}% bold math

\begin{document}
\title{Magnetic properties in the doped spin-1$/$2 honeycomb-lattice compound In$_3$Cu$_2$VO$_9$}

\author{Y. J. Yan$^1$, Z. Y. Li$^1$, T. Zhang$^1$, X. G. Luo$^1$, G. J. Ye$^1$, Z. J. Xiang$^1$, P. Cheng$^1$, Liang-Jian Zou$^2$}
\author{X. H. Chen$^1$}
\altaffiliation{Corresponding author} \email{chenxh@ustc.edu.cn}
\affiliation{1 Hefei National Laboratory for Physical Science at
Microscale and Department of Physics, University of Science and
Technology of China, Hefei, Anhui 230026, People's Republic of
China}
\affiliation{2 Institute of Solid State Physics, Chinese
Academy of Sciences, Hefei, Anhui 230031, People's Republic of
China}

\date{\today}% It is always \today, today,
             %  but any date may be explicitly specified

\begin{abstract}
We report the magnetic properties in the Co- and Zn-doped spin-1$/$2
honeycomb-lattice compound In$_3$Cu$_2$VO$_9$. The magnetic
susceptibility and specific heat experiments show no long-range
ordering down to 2 K in In$_3$Cu$_2$VO$_9$. In the low temperature
range, approximately T$^{2}$-dependent magnetic specific heat and
linearly T-dependent spin susceptibility were observed, suggesting a
spin liquid candidate with a S = 1$/$2 honeycomb lattice. When
Cu$^{2+}$ ions are partially substituted by Co$^{2+}$ ions, both
impurity potential scattering and magnetic impurity scattering
induced by magnetic Co$^{2+}$ ions break the homogenous spin-singlet
spin liquid state, and lead to an antiferromagnetic(AFM) long-range
correlation. While replacing Cu$^{2+}$ with nonmagnetic Zn$^{2+}$
ions, the frustration and antiferromagnetic correlation between
Cu$^{2+}$ ions is weakened, leading to breakage of a spin liquid
state and suppression of the low-dimensional AFM.
\end{abstract}

% PACS, the Physics and Astronomy
                             % Classification Scheme.
%\keywords{Suggested keywords}%Use showkeys class option if keyword
                              %display desired
\maketitle

\section{\label{sec:level1}Introduction}

Dimensionality and the spin magnitude S play important roles in the
physical properties of interacting systems because the quantum
fluctuation is affected significantly by them. The quantum
fluctuation enhanced both by geometric frustration and by spin
frustration may destroy the antiferromagnets(AFM) order and yield a
rich variety of ground states, novel excitations and exotic
behaviors that currently attract much attention. The magnetic
properties of a solid reflect the arrangement of the magnetic ions
in its crystal structure. Low-dimensional antiferromagnets exhibit a
variety of ground states depending on the spin number and the spin
configuration \cite{Ramirez}. In the first few years following the
discovery of high temperature superconductivity by Bednorz and
M\"{o}ller \cite{J. G. Bednorz}, S=1$/$2 2D systems on a square
lattice have been in the center of attention. Since the discovery of
the honeycomb-lattice Heisenberg antiferromagnet (AF)
Bi$_3$Mn$_4$O$_{12}$(NO$_3$), it revealed a novel spin-liquid-like
behavior down to low temperature which is ascribed to the
frustration effect due to the competition between the AF nearest-
and next-nearest-neighbor interactions J$_1$ and J$_2$ \cite{O.
Smirnova, M. Matsuda, S. Okumura}, growing interest arises in the
low-dimensional magnets with honeycomb lattice. The honeycomb
lattice is a loosely-coupled lattice with the number of the
neatest-neighbor sites only three, it might be susceptible to the
fluctuation effect caused by frustration, and its ordering property
is of special interest. Experimentally there are only a few examples
of materials where the electron spins are located in a
two-dimensional honeycomb lattice. However, diverse phenomena, such
as spin-glass state \cite{M. Bieringer, N. Rogado}, spin-liquid
state \cite{O. Smirnova}, Kosterlitz-Thouless transitions \cite{H.
M. Ronnow, M. Heinrich}, or superconductivity \cite{S. Shamoto, R.
Weht}, have been reported.

Recently, a complex transition metal oxide, In$_3$Cu$_2$VO$_9$, was
suggested as a possible candidate for the realization of the S=1$/$2
honeycomb lattice \cite{V. Kataev}. In$_3$Cu$_2$VO$_9$ was
previously reported to crystallize in the hexagonal space-group
P6$_3$$/$mmc, consisting of alternating layers of [InO$_6$]
octahedra, and Cu$^{2+}$ and V$^{5+}$ ions in a trigonal-bipyramidal
coordination \cite{V. Kataev}. The Cu$^{2+}$ (3d$^{9}$, S=1$/$2)
ions were proposed to be arranged in a 2D network of hexagons with
the nonmagnetic V$^{5+}$ (3d$^{0}$) ions in the center of each
hexagon. A later structural neutron diffraction study revealed that
a structural $\{$V1Cu6$/$3$\}$ order in the hexagonal planes has a
finite correlation length $\xi$$_{st}$ $\sim$ 300 ${\AA}$ and that
these structural domains are randomly arranged along the c axis
\cite{A. Moller}. The static susceptibility $\chi$(T) shows a broad
maximum at T$_0$ $\sim$ 180 K, being a characteristic feature of a
low-dimensional antiferromagnet, and passes through a kink at T$_1$
$\sim$ 38 K followed by a peak at T$_2$ $\sim$ 28 K. The anomaly at
T$_2$ previously identified with the transition to the long range
ordered state \cite{V. Kataev} have been tentatively assigned to
glass-like order of unsaturated spins in domain boundaries by
M\"{o}ller {\it et al.} \cite{A. Moller}. The real origin of the
peak-like anomaly is under debate.

To obtain insights into the nature of the puzzle properties of spin
liquid, especially the ground state of In$_3$Cu$_2$VO$_9$, we have
investigated the magnetic properties of the Co- and Zn-doped
spin-1$/$2 honeycomb-lattice compound In$_3$Cu$_2$VO$_9$. We show
that the T-dependent magnetic susceptibility and specific heat in
undoped In$_3$Cu$_2$VO$_9$ resemble to that of a spin liquid. When
the Cu$^{2+}$ ions are partially substituted by Co$^{2+}$ ions, the
low-dimentional antiferromagnetism (AFM) at 180 K is quickly killed,
while a long-rang AFM transition can be observed in the temperature
range from 50 K to 80 K, depending on Co concentration. When doping
Zn into the Cu sites, the low-dimensional AFM and the anomaly at
T$_2$ in the magnetic susceptibility are gradually suppressed, and
completely disappear with heavily Zn doping level.

\section{Material Preparation and Methods}

In$_3$Cu$_2$VO$_9$ polycrystalline pellets were synthesized by a
conventional solid-state technique. The starting materials,
In$_2$O$_3$, CuO, and V$_2$O$_5$ in a molar ratio of 3$:$$4$:$1$
were thoroughly ground and pressed into pellets. They were then
heated at 1173 K in air for five days with several intermediate
grindings and pelleting. In$_3$Cu$_{2-x}$Co$_x$VO$_9$ or
In$_3$Cu$_{2-x}$Zn$_x$VO$_9$ was synthesized using In$_2$O$_3$, CuO,
V$_2$O$_5$, Co$_3$O$_4$ or ZnO as starting materials. The raw
materials were accurately weighed according to the stoichiometric
ratio of the chemical formulas, and then synthesized using the
similar procedure to In$_3$Cu$_2$VO$_9$.

The samples were characterized by X-Ray diffraction (XRD) using
Rigaku D/max-A X-Ray diffractometer with Cu K$_\alpha$ radiation in
the range of 10$^{\circ}$-70$^{\circ}$ with the step of
0.02$^{\circ}$ at room temperature. Sample purity was checked by
powder x-ray diffraction, which showed no impurity peaks except for
the sample In$_3$Cu$_{1.5}$Co$_{0.5}$VO$_9$ in which the trace of
$In_2O_3$ is observed. Magnetic susceptibility was measured using
Vibrating Sample Magnetometer (VSM). Specific-heat measurements were
carried out from 2 K to room temperature using Quantum Design
Physical Property Measurement System (PPMS).

\section{Results and Discussion}

\begin{figure}[htbp]
\includegraphics[width=13 cm]{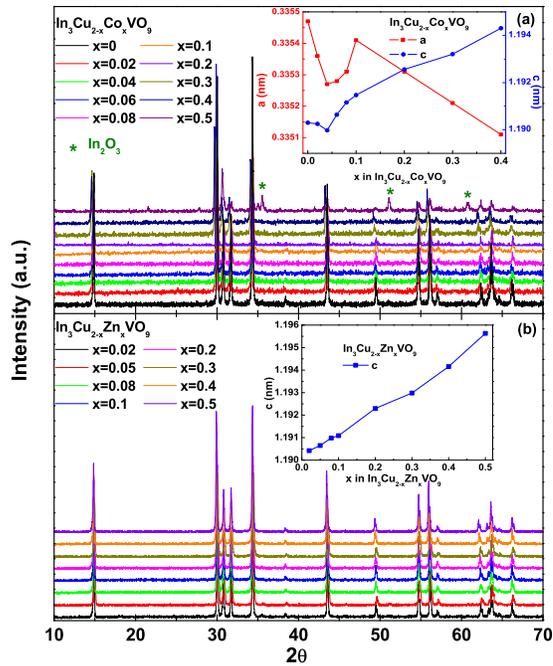}
\caption{(Color online) Powder x-ray diffraction patterns at room
temperature for polycrystalline samples, (a):
In$_3$Cu$_{2-x}$Co$_x$VO$_9$, (b): In$_3$Cu$_{2-x}$Zn$_x$VO$_9$. The
peaks marked with stars reveal the existence of the impurity phase
In$_2$O$_3$. Insets are doping dependence of a and c-axis lattice
parameters.}\label{Fig:Fig1}
\end{figure}

Fig.1(a) and (b) show the powder x-ray diffraction patterns for
In$_3$Cu$_{2-x}$Co$_x$VO$_9$ and In$_3$Cu$_{2-x}$Zn$_x$VO$_9$
polycrystalline samples, respectively. It is found that all the
peaks in XRD diffraction pattern can be well indexed to a hexagonal
structure. As seen in the inset of Fig. 1(a), when Co doping
concentration is less than 0.1, the lattice parameters for a- and
c-direction decrease firstly, reach minima at x=0.04, and then
increase with further Co doping. When Co doping concentration is
more than 0.1, the lattice parameter of a-direction decreases
monotonously with increasing Co doping concentration while that of
c-direction increases monotonously. In the
In$_3$Cu$_{2-x}$Zn$_x$VO$_9$ system, the lattice parameter for
c-direction increases with Zn doping concentration, while the change
of the lattice parameter for a-direction is not obvious.

\begin{figure}[htbp]
\includegraphics[width=15 cm]{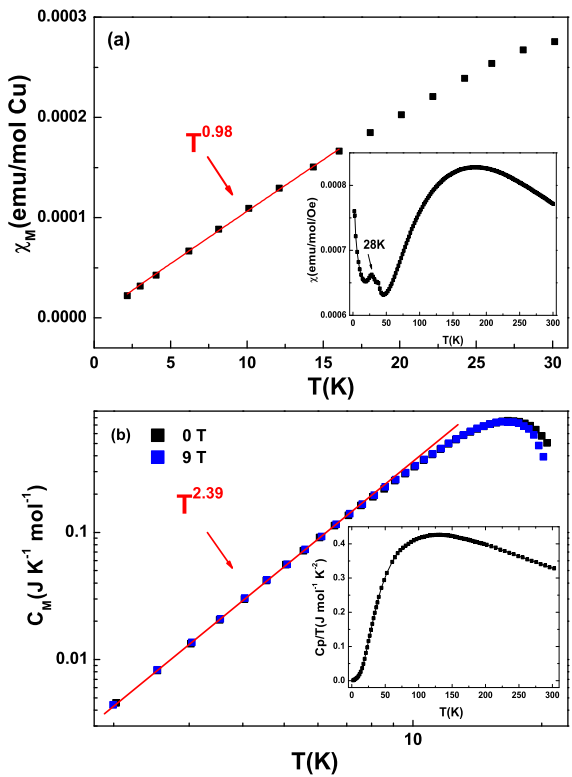}
\caption{(Color online)(a): Temperature dependence of spin
susceptibility of Cu$^{2+}$ ions in In$_3$Cu$_2$VO$_9$ parent
compound below 30 K under 1 T.  The inset of Fig.2(a) shows the
static magnetic susceptibility of In$_3$Cu$_{2}$VO$_9$ in the
temperature range from 2 K to 300 K under 1 T. (b): Temperature
dependence of magnetic specific heat C$_M$ of Cu$^{2+}$ ions. The
inset of Fig.2(b) shows temperature dependence of specific heat
divided by T for In$_3$Cu$_2$VO$_9$. The red solid lines are fitting
curve as described in the text.}\label{Fig:Fig1}
\end{figure}

The static magnetic susceptibility $\chi$(T) for
In$_3$Cu$_{2}$VO$_9$ was measured in a magnetic field of 1 T. The
temperature dependence of $\chi$ for In$_3$Cu$_2$VO$_9$ reveals a
broad maximum at T$_0$ $\sim$ 180 K, being a characteristic feature
of a low-dimensional antiferromagnet, and passes through a kink at
T$_1$ $\sim$ 38 K followed by a peak at T$_2$ $\sim$ 28 K and shows
a Curie-like upturn at lower temperatures (as seen in the inset of
Fig. 2 (a)). Temperature dependence of specific heat divided by T
for In$_3$Cu$_2$VO$_9$ is shown in the inset of Fig. 2 (b), no
signature for a magnetic transition above 2 K has been observed.
These results are consistent with the previous reports\cite{Y.
Fujii, M. Yehia, A. Moller}. Since the experimental data of both the
specific heat and magnetic susceptibility do not exhibit long-range
AFM ordering, one naturally assumed that the ground state of
In$_{3}$Cu$_2$VO$_9$ might be a gapless spin liquid phase, and a
long-time searched exotic quantum phase is performed both in theory
and in experiment. In Fig. 2, the spin susceptibility of Cu$^{2+}$
ions is obtained using the similar method as described in Ref. 12
(details given in Ref. 16). A nearly T-linear dependence of
$\chi$$_M$ can be observed by fitting the data between 2 K and 16 K
as is seen in Fig. 2 (a). The magnetic specific heat of Cu$^{2+}$
ions below 30 K is obtained by subtracting the phonon contributions
(details given in Ref. 17), which is shown in Fig. 2(b). Several
features have to be mentioned: (i) C$_M$ shows nearly no field
dependence with $\mu$$_0$H $\leq$ 9 T, and it is common in spin
liquid candidates \cite{H. D. Zhou}. This behavior has also been
observed for other spin liquid candidates, such as NiGa$_2$S$_4$
with Ni$^{2+}$(S=1) triangular lattice \cite{S. Nakatsuji} and
Na$_4$Ir$_3$O$_8$ with Ir$^{4+}$(S=1$/$2) hyperkagome lattice
\cite{Y. Okamoto}. (ii) with the log-log scale, between 2 and 7.5 K,
C$_M$ can be fit as C$_M$ = bT$^{\alpha}$ with b = 1.29 mJ K$^{-3}$
mol$^{-1}$ and $\alpha$ = 2.39. (iii) C$_M$ shows a broad hump
around 17 K. According to the early study \cite{Y. Ran}, a gapless
spin liquid phase in Kagome lattice may demonstrate unusual
temperature dependence behavior in the low-T range: the specific
heat is T$^{2}$-dependent, and the magnetic susceptibility linearly
increases with increasing temperature. Therefore, we find that our
experimental data on the T-dependent magnetic specific heat and spin
susceptibility almost obey these predictions for a spin liquid state
in a kagome lattice, in addition to small deviations which are
attributed to the contributions from magnetic impurities and fitting
error. It indicates that the ground state of In$_3$Cu$_2$VO$_9$
might be a spin liquid phase.

Theoretically, Sondhi {\it et al.} \cite{B. K. Clark} showed that in
a honeycomb Heisenberg model, when the next-nearest-neighbour spin
coupling is larger than 0.08 times and smaller than 0.3 times of the
nearest-neighbour spin coupling, strong spin frustration and spin
fluctuations can stabilize a spin liquid phase in a Heisenberg
insulator. In$_3$Cu$_2$VO$_9$ is an charge transfer insulator with
the energy gap of 1.6 eV, which is confirmed by the GGA+U simulation
done by Gou {\it et al.}\cite{Zou}. Well localized magnetic moments
in honeycomb copper spins with S=1$/$2 form a hexagonal net in the
Cu-V-O layer, showing that the low-energy magnetic properties of
In$_3$Cu$_2$VO$_9$ can be described by a honeycomb Heisenberg model.
In this material, though the direct next-nearest-neighbour (NNN)
spin coupling between copper spins is small, the indirect
next-nearest-neighbour can not be negligible. On the one hand, the
Cu-Cu superexchange coupling via oxygen ions can contribute a
fraction of NNN interaction; on the other hand, wide V 3d bands and
its hybridization with coppers may contribute another fraction of
NNN superexchange coupling. Hence, the NNN spin coupling and spin
fluctuation \cite{V. Kataev} may play important role to drive the
system from the N\'{e}el AFM ground state to the spin liquid one.

\begin{figure}[htbp]
\includegraphics[width=13 cm]{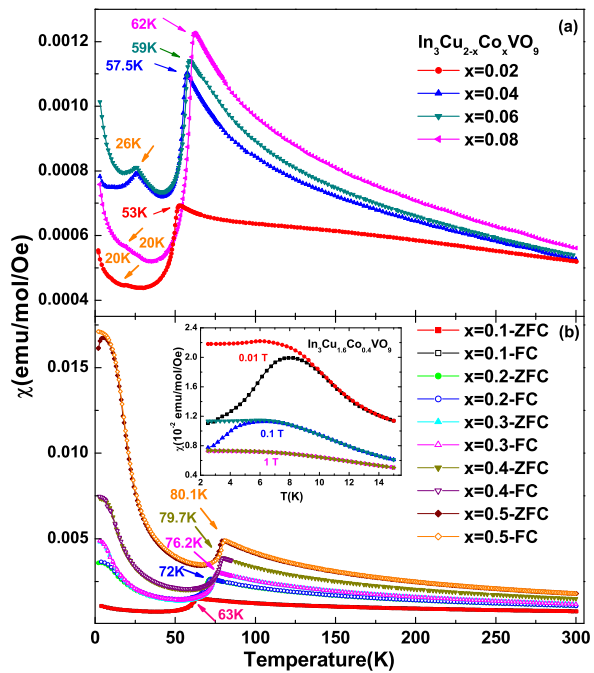}
\caption{(Color online) Temperature dependence of magnetic
susceptibility for In$_3$Cu$_{2-x}$Co$_x$VO$_9$ polycrystalline
samples in the temperature range from 2 K to 300 K under 1 T. Arrows
denote the occurrence temperature of the long range AFM ordering and
the peak-like anomaly. The inset shows the magnetic susceptibility
of In$_3$Cu$_{1.6}$Co$_{0.4}$VO$_9$ at low temperature under various
fields.}\label{Fig:Fig2}
\end{figure}

Temperature dependence of the static magnetic susceptibility for the
In$_3$Cu$_{2-x}$Co$_x$VO$_9$ samples is shown in Fig.3 (a) and (b).
The Weiss temperature is positive which is obtained by fitting the
magnetic susceptibility between 200 K and 300 K to the Curie-Weiss
law $\chi$=C$/$(T-$\theta$)+$\chi$$_0$ (C is curie constant,
$\theta$ is the Weiss temperature, and $\chi$$_0$ is the
temperature-independent term), indicating an antiferromagnetic
interaction between divalent transition metal ions (TM$^{2+}$). The
effective magnetic moment of the transition ions TM$^{2+}$ increases
from 1.29 $\mu$$_{B}$ for In$_3$Cu$_{1.96}$Co$_{0.04}$VO$_9$ to
2.32$\mu$$_{B}$ for In$_3$Cu$_{1.6}$Co$_{0.4}$VO$_9$. The spin
number of Co$^{2+}$ ions is 3$/$2. When the Cu$^{2+}$ ions are
partially substituted by Co$^{2+}$ ions, the low-dimensional
antiferromagnetism(AFM) at about 180 K is quickly destroyed, while a
long-rang AFM transition can be observed in the temperature range
from 50 K to 80 K (marked by arrows in Fig. 3), depending on Co
concentration. The transition temperature T$_N$ of the long-range
AFM order increases with Co concentration, tends a finite
temperature of about 80 K when x is large than 0.4. Meanwhile, one
notes that the magnitude of the magnetization increases with
increasing Co concentration due to the larger intrinsic magnetic
moment of Co$^{2+}$ than Cu$^{2+}$. When the Co doping level is less
than 0.08, there is a peak-like anomaly with a curie-like upturn
below 40 K. The anomaly is suppressed and disappears when Co doping
level is more than 0.08. As shown in Fig. 3(b), a trace of spin
glasslike contribution with T$_g$ $<$ 10K is observed. The
difference between zero-field cooling and field cooling
magnetization is suppressed with increasing field, and becomes
negligibly small at high fields above 1 T, as seen in the inset of
Fig. 3(b). This behavior is similar to that observed in the
hyperkagome lattice Na$_4$Ir$_3$O$_8$\cite{Y. Okamoto}.

\begin{figure}[htbp]
\includegraphics[width=9 cm]{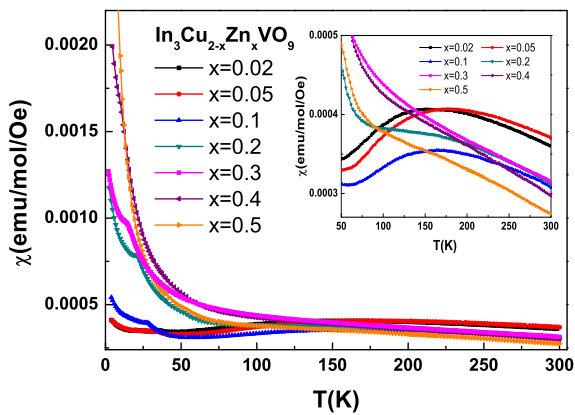}
\caption{(Color online) Temperature dependence of magnetic
susceptibility for In$_3$Cu$_{2-x}$Zn$_x$VO$_9$ polycrystalline
samples in the temperature range from 2 K to 300 K under 1 T. The
inset shows enlarged magnetic susceptibility $\chi$ (T) above 50
K.}\label{Fig:Fig3}
\end{figure}

In order to compare with Co doping, we tried to dope non-magnetic
Zn$^{2+}$ ions into Cu$^{2+}$ sites. Fig. 4 shows the temperature
dependence of static magnetic susceptibility of the
In$_3$Cu$_{2-x}$Zn$_x$VO$_9$ samples. By replacing Cu$^{2+}$ with
Zn$^{2+}$ ions, both the intensity and the temperature of the
peak-like anomaly are suppressed with increasing Zn concentration,
and the anomaly is killed when Zn concentration is more than 0.3. As
seen in the inset of Fig. 4, the broad hump at T$_0$ is notable when
Zn concentration x is less than 0.2, and tends to be suppressed with
higher Zn doping, and disappears when x $\geq$ 0.3. The Weiss
temperature and effective magnetic moment for x=0.3, 0.4, 0.5
samples are 19.6 K, 2.3 K, 1.2 K and 0.30 $\mu$$_{B}$, 0.29
$\mu$$_{B}$, 0.26 $\mu$$_{B}$, respectively. They are much smaller
than that of Co-doped samples. Therefore, with increasing Zn$^{2+}$
doping concentration, the frustration is suppressed gradually, and
the antiferromagnetic correlation strength between Cu$^{2+}$ ions is
weakened, leading to the breakage of a spin liquid state and
suppression of the low-dimensional AFM. Once x $\geq$ 0.3, the
strength of the correlation between Cu$^{2+}$ ions is not large
enough to form antiferromagnetic order, and a paramagnetic state is
observed.

\begin{figure}[htbp]
\includegraphics[width=9 cm]{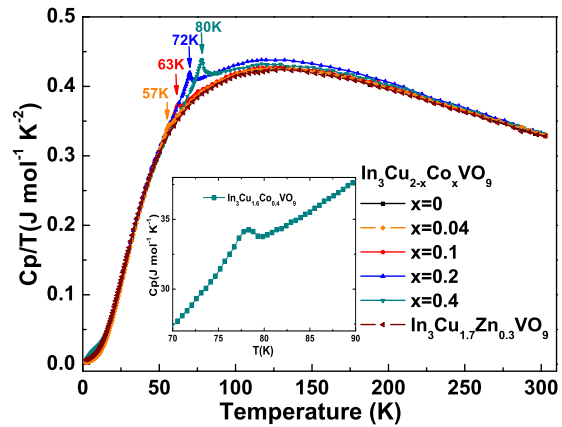}
\caption{(Color online) Temperature dependence of specific heat
divided by T for the samples of In$_3$Cu$_{2-x}$Co$_x$VO$_9$ and
In$_3$Cu$_{2-x}$Zn$_x$VO$_9$ systems. The inset shows the anomaly in
specific heat for In$_3$Cu$_{1.6}$Co$_{0.4}$VO$_9$ around the phase
transition.}\label{Fig:Fig4}
\end{figure}

Fig. 5 shows the specific heat divided by T for partial samples of
these two systems. A $\lambda$-shape-like jump is observed in all
the Co-doped samples ranging from 50 K to 80 K, which is ascribed to
long range AFM transition. For In$_3$Cu$_{1.7}$Zn$_{0.3}$VO$_{9}$
without low-dimensional AFM behavior, no anomaly is observed around
the temperature corresponding to the peak at about T$_2$ of the
magnetic susceptibility. Therefore, the peak-like anomaly of the
magnetic susceptibility is not associated with long-rang ordering.
It is inconsistent with the results reported by Yehia et al.
\cite{M. Yehia}, they claimed that strong experimental evidence for
the formation of the N\'{e}el type collinear AFM spin structure was
found in the S=1$/$2 honeycomb plane at temperatures below $\sim$ 20
K. In contrast, a neutron diffraction study revealed that a
structural $\{$V1Cu6$/$3$\}$ order in the hexagonal planes has a
finite correlation length $\xi$$_{st}$ $\sim$ 300 ${\AA}$ and these
structural domains are randomly arranged along the c axis \cite{A.
Moller}. The peak-like anomalies have been tentatively assigned to
glass-like order of unsaturated spins in domain boundaries by
M\"{o}ller {\it et al.}\cite{A. Moller}. Therefore, we consider the
peak-like anomaly of magnetic susceptibility is not associated with
long-rang ordering, and might arise from unsaturated spins in domain
boundaries.

After substituting Cu$^{2+}$ with Co$^{2+}$ in In$_3$Cu$_2$VO$_9$,
one expects that the location of Co$^{2+}$ ions is random.
Obviously, both impurity potential scattering and magnetic impurity
scattering induced by Co$^{2+}$ ions (S = 3$/$2) break the
homogenous spin-singlet spin liquid state, releasing the AFM
long-range correlation and leading to the AFM behaviors in specific
heat and magnetic susceptibility. Such a scenario can address
various T-dependent properties over wide doping range we observed in
experiments, for example the various AFM-paramagnetic (PM) phase
transition peak as observed in specific heat and magnetic
susceptibility in many doped samples. The antiferromagnetic
correlation length of magnetic TM$^{2+}$ ions increases with
increasing Co$^{2+}$ concentration, leading to higher AFM transition
temperature. Meanwhile, by replacing Cu$^{2+}$ with nonmagnetic
Zn$^{2+}$ ions which is also random located, the antiferromagnetic
correlation between Cu$^{2+}$ ions is destroyed, breaking the
homogenous spin-singlet spin liquid state and suppression of the
low-dimensional AFM.

\section{Conclusion}

In summary, we successfully synthesized Co- and Zn-doped spin-1$/$2
honeycomb-lattice compound In$_3$Cu$_{2-x}$Co$_x$VO$_9$ (0 $\leq$ x
$\leq$ 0.5) and In$_3$Cu$_{2-x}$Zn$_x$VO$_9$ (0 $\leq$ x $\leq$
0.5), and studied their magnetic properties. The magnetic
susceptibility and specific heat experiments of In$_3$Cu$_2$VO$_9$
show no long-range ordering down to 2 K. Approximately
T$^{2}$-dependent magnetic specific heat and linearly T-dependent
spin susceptibility in the low temperature range were observed in
In$_3$Cu$_2$VO$_9$, suggesting a spin liquid candidate with a S =
1$/$2 honeycomb lattice. When the Cu$^{2+}$ ions are partially
substituted by Co$^{2+}$ ions, the low-dimensional
antiferromagnetism(AFM) at about 180 K is quickly destroyed, while a
long-rang AFM transition is observed in the temperature range from
50 K to 80 K, depending on Co concentration. Both impurity potential
scattering and magnetic impurity scattering induced by magnetic
Co$^{2+}$ ions break the homogenous spin-singlet spin liquid state,
releasing the AFM long-range correlation. While replacing Cu$^{2+}$
with nonmagnetic Zn$^{2+}$ ions, the frustration and
antiferromagnetic correlation between Cu$^{2+}$ ions is weakened,
leading to breakage of a spin liquid state and suppression of the
low-dimensional AFM.

\vspace*{2mm} {\bf Acknowledgment:} This work is supported by the
Natural Science Foundation of China and by the Ministry of Science
and Technology of China, and by Chinese Academy of Sciences.


\begin{thebibliography}{}
\bibitem{16}
We analyze the static susceptibility as a sum of three contributions
$\chi$$_{stat}$ = $\chi$$_0$ + $\chi$$_{CW}$ + $\chi$$_M$, where
$\chi$$_0$ is the sum of the T-independent diamagnetic and Van-Vleck
susceptibility, $\chi$$_{CW}$ = C$_{def}$ $/$ (T + $\Theta$) is the
Curie-Weiss contribution owing to uncompensated spins at the defect
structural sites, and $\chi$$_M$ is the spin susceptibility of
Cu$^{2+}$ ions. In this paper, C$_{def}$ = 0.012 emu $/$ mol Cu,
$\Theta$ = 18 K and $\chi$$_0$ = 1.354 $\times$ 10$^{-4}$ emu $/$
mol Cu were used to obtain the spin susceptibility $\chi$$_M$ of
Cu$^{2+}$ ions.
\bibitem{17}
As reported in Ref. 13, three characteristic mode temperatures
$\Theta$$_E$ have been estimated from IR spectroscopy to be
$\approx$ 160, 350, and 710 K, corresponding to the cations In, Cu,
and V, respectively. In the low-temperature region, $\Theta$$_D$ =
268 K has been obtained by Debye model. The phonon contribution to
the specific heat is obtained using Einstein formula by combining
the Debye and Einstein contributions using the following weights
C$_{phon}$ = C$_{\Theta_D}$ + 0.333 C$_{\Theta_{E1}}$ + 1.334
C$_{\Theta_{E2}}$ + 2.333 C$_{\Theta_{E3}}$.
\end{thebibliography}
\end{document}